\documentclass[aps,prl,twocolumn,showpacs,superscriptaddress,
preprintnumbers,amsmath,amssymb]{revtex4}

\usepackage{graphicx}
\usepackage{dcolumn}
\usepackage{bm}
\usepackage{ifthen}
\usepackage{booktabs}
\def\mc{\multicolumn}

\begin{document}

\title{Origin of ferroelectricity in high $T_c$ magnetic ferroelectric CuO}
\author{Guangxi Jin}
\affiliation{Key Laboratory of Quantum Information, University of
Science and Technology of China, Hefei, 230026, People's Republic of
China}

\author{Kun Cao}
\affiliation{Key Laboratory of Quantum Information, University of
Science and Technology of China, Hefei, 230026, People's Republic of
China}

\author{Guang-Can Guo}
\affiliation{Key Laboratory of Quantum Information, University of
Science and Technology of China, Hefei, 230026, People's Republic of
China}

\author{Lixin He \footnote{Email address: helx@ustc.edu.cn}}
\affiliation{Key Laboratory of Quantum Information, University of
Science and Technology of China, Hefei, 230026, People's Republic of
China}

\date{\today }

\begin{abstract}
``Magnetic ferroelectric'' has been found in a wide range of spiral magnets.
However, these materials all suffer from low critical temperatures, which are usually
below 40 K, due to strong spin frustration. Recently, CuO has been
found to be multiferroic at much higher ordering temperature ($\sim$ 230K).
To clarify the origin of the high ordering temperature in CuO,
we investigate the structural, electronic and magnetic properties of CuO via
first-principles methods.
We find that CuO has very special nearly commensurate spiral magnetic structure,
which is stabilized via the Dzyaloshinskii-Moriya interaction. The spin
frustration in CuO is relatively weak, which is one of the main reasons that
the compound have high ordering temperature.
We propose that high $T_c$ magnetic ferroelectric materials can be found in
double sublattices of magnetic structures similar to
that of CuO.
\end{abstract}



\maketitle


Magnetic ferroelectric materials in which ferroelectricity is
induced by magnetic ordering, have attracted intensive interests
\cite{fiebig05,cheong07}.
The strong magnetoelectric (ME) coupling in these materials opens up a new path to
the design of multifunctional devices that allow
the control of charges by the application of magnetic fields or spins by applying
voltages.
So far, almost all magnetic ferroelectric materials are strongly frustrated magnets\cite{cheong07}.
Frustrated magnets have very low ordering temperatures  ($\sim$ 30 - 40 K),
several times smaller than the temperatures expected from their spin interaction strengths.
Low critical temperature is one of the major factors that limit
the applications of these important materials.
Therefore a new mechanism that allows high temperature magnetic ferroelectric materials
is critical.

\begin{figure}
\centering
\includegraphics[width=3.0in]{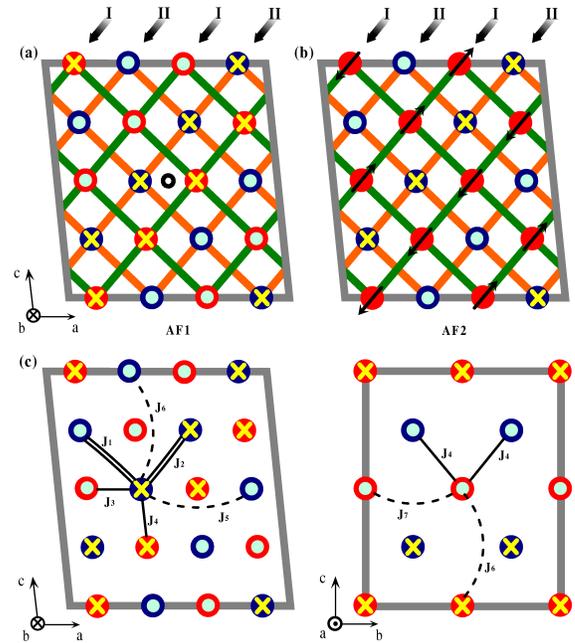}
\caption{ Schematic sketch of the magnetic structures of (a) the collinear AFM1
  phase, (b) the noncollinear AFM2 phase.  The black arrows, yellow crosses, and
  blue circles denote the spin directions associated with Cu ions.
  The black circle in (a) indicates an inversion center.
 (c) A sketch of superexchange interactions $J_1$ to $J_7$.
  The single lines, double lines, and dashed lines represent the three types of
  exchange interactions between Cu ions.}
\label{fig:spin}
\end{figure}

Recently, CuO was found to be multiferroic at $T_c$=230 K, which is
much higher than the critical temperatures of all other magnetic ferroelectric materials \cite{kimura08}.
However, the mechanism for the high ordering temperature
were not clear.
So far, CuO is the only binary compound that has been found to be multiferroic \cite{kimura08}.
CuO undergoes two successsive magnetic phase transitions upon cooling from room
temperature to near zero temperature.
Neutron scattering experiments \cite{yang89} show that
below $T_{N1}$=213 K, the spin structure is collinear antiferromagnetic (AFM1)
[see Fig. \ref{fig:spin}(a)].
Between $T_{N1}$ and $T_{N2}$=230 K, the spin structure becomes
non-collinear and slightly incommensurate (AFM2) [see Fig. \ref{fig:spin}(b)],
with a modulation vector of ${\bf Q}$= (0.006, 0, 0.017).
Remarkably, an electric polarization of 160 $\mu$C$\cdot m^{-2}$,
which can be reversed by applying an
electric field of about 55 kV/m, develops in the AFM2 phase.
The electric polarization was attributed to the spiral spin
structure \cite{kimura08,mostovoy06}
which was assumed to
  result from spin frustration and whereas the high ordering temperature is believed to
  come from the strong exchange interactions \cite{kimura08}.

To clarify the mechanism behind its high ordering temperature and the
origin of its ferroelectricity,
we carry out first-principles studies of the multiferroism
of CuO.
We find that CuO has very special nearly commensurate spiral magnetic structure,
which is stabilized via the Dzyaloshinskii-Moriya interaction
\cite{dzyaloshinskii64,moriya60}. The spin
frustration in CuO is relatively weak, which is responsible for its high ordering
temperature. The work suggest that high $T_c$
magnetic ferroelectric materials can be found in materials which have
double sublattices of magnetic structures similar to that of CuO.

The crystal structure of CuO is monoclinic containing four chemical formulas per unit cell.
The AFM1 spin structure is composed of two antiferromagnetic (AFM)
  spin sublattices, in which Cu ions have the same $b$ values in each sublattice.
The spin chains along the [10$\bar{1}$] direction are antiferromagnetic and are labeled chain
I and chain II for the two sublattices,
whereas the chains along the [101] direction are ferromagnetic.
In the AFM1 phase, all spins are aligned in the $b$ direction, whereas
in the AFM2 phase, chain II rotates perpendicularly to chain I.
To accommodate the spin structures, we use a 2$\times$1$\times$2 CuO supercell, which contains 32 atoms.
For the AFM2 structure, we neglect the small incommensurate component of the
spin structure [i.e., set ${\bf Q}$= (0, 0, 0)] and rotate the spin
directions of chain II 90$^\circ$, so that it lies in the $ac$ plane. In the
AFM2 phase, the spins form cycloidal spirals along the $a$ and $c$ axis.
The incommensurate component of the magnetic modulation vector \textbf{Q} is extremely
small, and it should not affect the calculated electric polarization because
$\textbf{P} \propto \textbf{S}_i \times \textbf{S}_j$ \cite{mostovoy06}.

\begin{table}
\caption{Calculated lattice parameters of CuO compared with the experimental
  results obtained at room temperature.
  The theoretical result is calculated using collinear
  antiferromagnetic (AFM1) spin structure. The calculated
  crystal structure is symmertized accroding to the C2/c symmetry.}
\label{tab:structure}
\begin{tabular}{cccc}
\hline \hline
lattice constant  &    Expt. (Refs.\cite{yang88})   &  LSDA+U \\
\hline
 $a$ (\AA)             & 4.6837                                      & 4.5914  \\
 $b$ (\AA)             & 3.4226                                      & 3.3277  \\
 $c$ (\AA)             & 5.1288                                      & 5.0268  \\
 O ($b$)               & 0.4162                                      & 0.4110  \\
$\beta$                & 99.54$^0$                                   & 100.025$^0$ \\
\hline 
\hline
\end{tabular}
\end{table}

We perform ab initio calculations on CuO, based on the density functional
theory (DFT) within the spin-polarized local density approximation (LSDA)
implemented in the Vienna ab initio simulations package (VASP) \cite{kresse93,kresse96}.
The on-site Coulomb interactions $U$ are included for Cu ions in the rotationally
invariant scheme introduced by Liechtenstein et al. \cite{liechtenstein95}.
The spin-orbit coupling is taken into account in the calculation unless otherwise
noticed.
We have tried several $U$ values.
We present here mainly
the results for $U$=7.5 eV and $J$=0.98 eV, which are typical values for Cu \cite{anisimov91}.
Projector-augmented-wave (PAW) pseudopotentials \cite{blochl94} with a
500 eV plane-wave cutoff are used.
A 2$\times$4$\times$2 $k$-points mesh converges the results very
well. We relax the structure until the changes of total
energy in the self-consistent calculations are less than
10$^{-5}$ eV and the remaining forces are less than 1 meV/\AA.

To determine the crystal structure under the AFM1 spin configuration,
we relax the structure starting from experimentally determined structures.
The room temperature crystal structure is monoclinic and of space group C2/c, with
inversion symmetry. However, the magnetic structure of the AFM1 phase has
only $P2_1/c$ symmetry. Therefore, after relaxation, the crystal structure is also
reduced to $P2_1/c$ symmetry, because of the ``exchange striction''
effects. To get an idea of the amplitude of the lattice distortion, we symmetrize
the crystal structure with $P2_1/c$ symmetry to a crystal structure with C2/c
symmetry, following Ref.\cite{wang08}. We find that the Cu and O ions deviate
from their high symmetry sites by about 10$^{-3}$ \AA. 
This distorted structure preserves the inversion symmetry; 
therefore, it has no net polarization. The inversion center is shown in Fig. \ref{fig:spin}(a).
An inversion operation about the inversion center changes spin chain I to chain II.
The obtained structural parameters are shown in Table \ref{tab:structure},
and the calculated structural parameters are in good agreement with the experiments.

\begin{table}
\caption{Comparison of the total energies (in eV) of phases AFM1 and AFM2,
with/without spin-orbit coupling (soc) and
with/without structure relaxation (relax).}
\centering 
\begin{tabular}{lcc} 
\hline\hline 
     & AFM1      & AFM2 \\
\hline 
 no soc + relax         & 0.58643    &  0.59588  \\
 no soc + no relax      & 0.59481    &  0.59588  \\
 soc + relax            & 0          &  0.01083  \\
 soc + no relax         & 0.00822    &  0.01087  \\
\hline 
\hline     
\end{tabular}
\label{tab:totalenergy}
\end{table}

We calculate the density of states and the band gap of CuO of the AFM1 phase.
If no on-site Coulomb interaction is presented for Cu, the system is metallic.
For $U$=7.5 eV, the calculated band gap is 2.01 eV, which is close to experimental
value that ranges from 1.35 to 1.6 eV \cite{koffyberg82, ghijsen88,marabelli95}.
The calculated magnetic moments for the Cu ion is
0.70$\mu_B$, where $\mu_B$ is the Bohr magneton,
also in good agreement with experimental value 0.65-0.68$\mu_B$ \cite{forsyth88,yang88}.
The oxygen ions also have significant induced magnetic moments for about 0.142 $\mu_B$.

To get the crystal structure under the AFM2 spin consfiguration, we
constrain the spin orientations to the AFM2 configuration and
relax the structure again. The relaxation also starts from
the experimental high symmetry crystal structure.
We fix the lattice vectors $a$, $b$ and $c$
to those obtained from the AFM1 structure.
We compare the total energies of the AFM1 and AFM2 phases, shown in Table
\ref{tab:totalenergy}. After relaxation,
the total energy of the AFM2 phase is about 0.33 meV per atom higher than that of
the AFM1 phase, which is consistent with experiments that AFM2 phase appears at a higher
temperature than the AFM1 phase \cite{yang89, brown91, ain92}.
Before relaxation, the total energy of the AFM2 phase is only about 0.08 meV per atom higher than
the AFM1 phase, mostly due to spin anisotropy energy. This can be seen from
that if spin-orbit coupling is turned off, the energy difference between
the two phases reduces to 0.03 meV per atom. This indicates that the
spiral configuration is not stable without spin-orbit interactions, i.e., the
system is almost invariant under a general common rotation.
The remaining 0.25 meV per atom is due to the fact that the ionic distortion
in the AFM1 phase is much larger than that of the
AFM2 phase, which will be shown below.
The band gap of the AFM2 phase is slightly smaller than that of the AFM1 phase. For
$U$=7.5 eV, the band gap of the AFM2 phase is about 1.78 eV. The local magnetic
moment of the Cu and the oxygen ions are also slightly smaller than those of the AFM1
phase, being 0.65 $\mu_B$ and 0.107 $\mu_B$, respectively.

\begin{table}
\caption{Nonequvalent atomic positions of the AFM2 phase in the
  2$\times$1$\times$2 CuO super cell.  $\delta a$, $\delta b$, and $\delta c$
  are the deviation of atomic positions from high symmetry (C2/c) positions.}
\begin{tabular}{lccccccccc}
\hline \hline
                          & \mc{3}{c} {AFM2}          &\mc{3}{c}{distortion (10$^{-6}$)}  \\
atom         &  $a$       & $b$        &  $c$        & $\delta a$ &$\delta b$  &$\delta c$ \\
\hline
Cu$_1$       & 0.124988    & 0.249983  & 0.000007    & -12.0    & -17.2      &  6.9  \\
Cu$_2$       & 0.374985    & 0.749978  & 0.000005    & -15.0    & -21.9      &  5.4  \\
Cu$_3$       & 0.375012    & 0.249983  & 0.249993    &  12.0    & -17.2      & -6.9  \\
Cu$_4$       & 0.125015    & 0.749978  & 0.249995    &  15.0    & -21.9      & -5.4  \\
O$_1$        & 0           & 0.410727  & 0.125       & 0        &  16.5       & 0    \\
O$_2$        & 0           & 0.589308  & 0.375       & 0        &  17.9       & 0    \\
O$_3$        & 0.25        & 0.910733  & 0.125       & 0        &  22.7       & 0    \\
O$_4$        & 0.25        & 0.089311  & 0.375       & 0        &  21.2       & 0    \\
\hline 
\hline
\end{tabular}
\label{tab:AFM2}
\end{table}

A symmetry analysis shows that the AFM2 spin structure does NOT have
inversion symmetry due to the rotation of spin chain II, which can
be easily seen in Fig. \ref{fig:spin}(b). The crystal
structure is distorted by the Dzyaloshinskii-Moriya (DM) interaction
\cite{dzyaloshinskii64,moriya60,sergienko06}, which breaks the inversion symmetry. The structural
parameters of the AFM2 phase are shown in Table \ref{tab:AFM2}.
Compared with the high-symmetry structure, all the oxygen ions are shifted
in the +b direction for about 7$\times$10$^{-5}$ \AA, whereas all the
Cu ions are shifted in the -b direction by a similar amount. The
ionic distortion in the AFM2 phase is approximately 2 orders of
magnitudes smaller than that driven by the ``exchange striction'' 
effects in the AFM1 phase.

Next, we calculate the electric polarization using the Berry-phase theory of
polarization\cite{king-smith93}.
The calculated total polarization is about 90 $\mu$C m$^{-2}$ in the -$b$
direction, which is somewhat smaller than the experimental value of 160 $\mu$C$\cdot$
m$^{-2}$ along the $b$ axis. 
The agreement between theoretical calculations and experimental values is
reasonable, given that the current functionals are not adequate to treat the
subtle correlation effects in 
magnetic ferroelectric materials \cite{moskvin08}.

In magnetic ferroelectric materials, the electric polarization can be either purely
electronic or ionic. To separate the two contributions,
we calculate the electric polarization in the AFM2 spin configuration,
using the symmetrized crystal structure with C2/c symmetry.
We obtain an electric polarization about
38 $\mu$C $\cdot$ m$^{-2}$ along the $b$-axis, which is the pure electronic
contribution to the polarization. The ionic contribution is then the remaining -128
$\mu$Cm$^{-2}$. Therefore, the pure electronic contribution and ionic
contribution are of the same order of magnitude, but of opposite sign. We
also calculate the electric polarization using different on-site Coulomb $U$
values for the Cu ions. We find that the polarization is very sensitive to the $U$
values, because the electronic contribution and ionic
contribution to the polarization have opposite signs.
For example, if $U$=5 eV is used, the polarization is reduced to
about 30 $\mu$C $\cdot$ m$^{-2}$. We note that the Coulomb
$U$ is empirically chosen, which is somewhat unsatisfactory. However, this is the
best calculations one can do presently. In the future,
parameter free, and yet very computationally demanding methods,
such as hybrid functional DFT \cite{perdew96b} can be used to solve this problem.

We repeat the above calculations without
including spin-orbital coupling. We find that the lattice distortion is nearly absent
after turning off the spin-orbit interaction for the AFM2 phase. At the same
time the electric polarization including both the purely electric and the
ionic parts is nearly eliminated. This confirms that the spin-orbit
interaction is essential to the lattice distortion and electric polarization
in this material.

In magnetic ferroelectric materials, the transition temperatures are predominantly
determined by the magnetic exchange interactions \cite{kimura08}.
We extract the superexchange interactions $J$s of CuO using a Heisenberg model
$H=-\sum_{ij} J_{ij} {\bf S}_i \cdot {\bf S}_j - \sum_i(\textbf{K} \cdot \textbf{S}_i)^2$
from calculated total energies of the different spin configurations in the
symmetrized C2/c crystal structure with spin-orbit coupling. \textbf{K} is the
anisotropic energy due to spin-orbit coupling.
There are 7 $J$s in total, which are shown in Fig. \ref{fig:spin}(c).
Among them, $J_1$ is the exchange interactions between nearest-neighbor Cu atoms along
the [10$\bar{1}$] direction, 
$J_2$ is the interaction between nearest-neighbor Cu atoms along the [101]
direction, $J_7$ is the interaction between the nearest-neighbor spins of the
same sublattice along the b direction, and $J_3$ and $J_4$ are the
intersublattice exchange interactions. The fitted $J_1$= -51 meV is in a good
agreement with $J$=67$\pm$20 retrieved from neutron scattering experiments
\cite{yang89}. The fitted value of $J_2$=8.6 meV, is only about 1/6 of
$|J_1|$, consistent with quasi-1D model \cite{shimizu03}, and the fitted value
of $J_7$=9.87 meV. The fitted intersublattice coupling values $J_3$=4.9 meV
and $J_4$=7 meV are weak, because the Cu-O-Cu bond angles are close to
90$^\circ$ for these two $J$s \cite{shimizu03}.In AFM1 and AFM2 phases,
symmetry causes additional mutual cancelation of $J_3$ and $J_4$; therefore,
the energy cost of chain II rotation is low, as was discussed in the previous
paragraph.
We also calculate the next-nearest-neighbor (NNN) interactions $J_5$,
$J_6$. We find that $J_5$ and
$J_6$ are very asymmetric. $J_5$ is significant with a value of -12 meV, whereas $J_6$
is only 2.1 meV. These values agree well with those given in
Ref. \cite{filippetti05}.

The intrasublattice interactions $J_1$, $J_2$, and
\textbf{$J_7$} basically determine the ground state spin structure AFM1.
The next-nearest-neighbor interaction $J_5$
further favors the antiferromagnetic spin chain along the [10$\bar{1}$]
direction, whereas $J_6$ only adds a small frustration to this configuration.
The major competing interactions are those of the intersublattice
interactions $J_3$ and $J_4$.
The weak incommensurateness of the spin spiral caused by frustrated exchange
interactions $J_3$, \textbf{$J_4$} is consistent with that in this material
the spin competition is small.
We calculate the ordering temperature by forcing all exchange interactions to
be ferromagnetic, and get $T_c$= 311 K by a Monte Carlo simulation.
This temperature is only about 1.5 times greater than the $T_c$ of
the AFM1 phase from simulations
(In RMn$_2$O$_5$ \cite{cao09}, the ratio is approximately 3 - 4.),
which also indicate that the spin frustration is weak in CuO.
The lack of strong competing interactions in this compound might explain the
high spin-ordering temperature of CuO.

To summarize, we have investigated the structural, electronic and magnetic
properties of CuO.
We show that CuO has very special nearly commensurate spiral magnetic structure,
which is stabilized by the Dzyaloshinskii-Moriya interaction. The spin
frustration in CuO is relatively weak, which is one of the main reasons that it
has much higher ordering temperature than other magnetic ferroelectric materials.
We propose that high $T_c$ magnetic ferroelectric can be found in materials
which have double sublattices of magnetic structures similar to
that of CuO.

LH acknowledges the support from the Chinese National
Fundamental Research Program 2011CB921200, the Innovation
funds and ``Hundreds of Talents'' program from Chinese Academy of
Sciences.


\end{document}